\documentclass[journal=jctcce,manuscript=article]{achemso}
\usepackage{amssymb}
\usepackage{siunitx}
\usepackage{amsmath}
\usepackage{xcolor}
\usepackage[version=3]{mhchem} 

\usepackage{epsfig}
\usepackage[T1]{fontenc}
\setkeys{acs}{articletitle = true}

\author{Sara Roosta}
\affiliation{Institute of Physical Chemistry (IPC), Karlsruhe Institute of Technology, 76131 Karlsruhe, Germany}

\author{Marcus Elstner}
\affiliation{Institute of Physical Chemistry (IPC), Karlsruhe Institute of Technology, 76131 Karlsruhe, Germany}
\alsoaffiliation{Institute of Biological Interfaces (IBG2), Karlsruhe Institute of Technology, 76131 Karlsruhe, Germany}

\author{Weiwei Xie}
\affiliation{Frontiers Science Center for New Organic Matter, Haihe Laboratory of Sustainable Chemical Transformations,
Key Laboratory of Advanced Energy Materials Chemistry (Ministry of Education), State Key Laboratory of Advanced Chemical Power Sources,
College of Chemistry, Nankai University, Tianjin, 300071, China}
\email{xieweiwei@nankai.edu.cn}

\title{Effect of Halogen Substituents on Charge Transport Properties of n-type Organic Semiconductors:
A Theoretical Study}


\begin{document}

\begin{abstract}
Organic semiconductors (OSCs) have received much attention as promising materials for
electronic devices. In this study, we investigate the impact of halogen groups on the charge
transport properties of n-type OSC-6,13 bis ((triisopropylsilyl) ethynyl)-5,7,12,14-tetraazapentacene (TIPS-TAP).
The computed mobilities for TAPs substituted with F and Cl exhibit excellent agreement with the experimental values,
while the simulation overestimates the electron mobility for TAP. Interestingly, the mobility of TIPS-TAP-4F is significantly
lower than that of TIPS-TAP-4Cl/Br, despite their similar packing structures. This discrepancy can be attributed to the
strong electron-withdrawing effect of fluoride, leading to reduced electron transfer integrals and increased reorganization energy.
While molecular packing is widely accepted as a dominant factor in charge transport in OSCs, our study highlights the essential role of electronic effects in OSC charge transport.
This study provides new insights into the understanding of the charge transport mechanism in OSCs.

\end{abstract}

\maketitle

\section{Introduction}
Organic semiconductors (OSCs) \cite{forrest2004path,anikeeva2015restoring,minemawari2011inkjet,sun2012solution,li2023recent,cao2021pillararene}
are a class of material that have attracted great attention in recent
years and have been widely used for electronic and optoelectronic devices, such as organic light emitting diodes,
organic ﬁeld eﬀect transistors, and organic photovoltaics. Compared with inorganic 
semiconductors, these materials have outstanding properties due to their unique advantages of low
processing temperature, high quantum yields, low fabrication cost and good mechanical flexibility .
However, the weak van der Waals interactions in OSCs leads to weak couplings and  large dynamic ﬂuctuations,
which hamper the charge transport in OSCs.

Organic semiconductors can be broadly classified into two types based on their charge transport behavior: p-type and n-type.
P-type organic semiconductors are materials that exhibit a predominant positive charge carrier (hole) transport, while
n-type organic semiconductors exhibit a predominant negative charge carrier (electron) transport. Charge carrier mobility
is one of the key quantities to measure the performance of OSC devices. So far, several herringbone-stacked p-type OSCs have been found to exhibit
high hole mobilities exceeding 10 cm$^2$~V$^{-1}$~s$^{-1}$\cite{podzorov2005hall,kang2011alkylated,park2015dibenzothiopheno,liu2015thin,tsutsui2016unraveling,okamoto2020bent,yu2022mixed},
such as rubrene\cite{podzorov2005hall}, [1]benzothieno[3,2-b][1]benzothiophene  (BTBT) \cite{tsutsui2016unraveling}, 2, 6-diphenyl anthracene \cite{liu2015thin},
and recently developed bent-shaped molecules \cite{okamoto2020bent}. However,  there are relatively fewer literature reports on
n-type organic semiconductors (OSCs) exhibiting electron mobilities exceeding 10 cm$^2$~V$^{-1}$~s$^{-1}$ \cite{li2012high,dou2015fine,chu2018halogenated}.
To the best of our knowledge, the current record for the reported electron mobility is  27.8 cm$^2$~V$^{-1}$~s$^{-1}$, achieved using
chlorine-substituted 6,13-bis((triisopropylsilyl)ethynyl)-5,7,12,14-tetraazapentacene (TIPS-TAP) by Miao and et al\cite{chu2018halogenated}.
Chlorine substitution in TIPS-TAP introduces electron-withdrawing properties, which lowers the energy of the lowest unoccupied molecular
orbital(LUMO) and thus leads to a higher charge transfer rate. However, this work also shows that the substitution of a stronger electron-withdrawing group,
such as fluorine, results in the much lower electron mobility (6.6 cm$^2$~V$^{-1}$~s$^{-1}$) compared to TIPS-TAP-4Cl. The reason is still unknown.

In this study, we perform nonadiabatic molecular dynamics simulations using the fewest switched surface hopping (FSSH) method \cite{tully1990} to
study charge transport in a series of halogen-substituted TIPS-TAP. The FSSH method is the most widely used nonadiabatic dynamics method
for the simulation of charge transport process \cite{troisi2011dynamic,giannini2019quantum,roosta2022efficient}. In a previous study  \cite{roosta2022efficient},
we have shown that the FSSH method using Boltzmann correction (BC) for velocity adjustment and implicit charge relaxation (IR) approximation can reasonably reproduce the
experimental charge mobilities for a number of representative organic materials. For halogen-substituted TIPS-TAP, the electron mobilities obtained by the FSSH-IR-BC method
are in good agreement with experimental values. Moreover, it is found that despite the similar packing structures for all halogen-substituted TIPS-TAPs,
the strongest electron-withdrawing effect of fluoride lead to lowest electron transfer integrals and largest reorganization energy, resulting in the
lowest charge mobilities.

\section{Theoretical Methods}
\subsection{Quantum chemical calculations and the fragment orbital scheme}
The molecular complex is divided into two regions, the quantum chemical (QM) region, where
the excess charge carrier propagates and the remainder of the complex which is treated as
the molecular mechanics (MM) region. The energy of the QM region containing no excess
charge is approximated with MM, and is included in the total MM energy $E^\mathrm{tot}_\mathrm{MM}$
MM \cite{heck2015multi}. The total energy of a molecule in the QM zone with a electron can be approximated as

\begin{equation}\label{etot1}
E^-\approx E^\mathrm{tot}_\mathrm{MM} - \langle \Psi | H[\rho_0] |\Psi \rangle
+ \Delta E_\mathrm{QM/MM} \;,
\end{equation}
where $| \Psi \rangle $ is the electron wave function, $H[\rho_0]$ is the Kohn--Sham
Hamiltonian of the neutral system, and $\Delta E_\mathrm{QM/MM}$ is the interaction
energy between the QM and MM subsystems. Since organic semiconductors consist of weakly bound molecules\cite{heck2015multi},
we use a simple coarse-graining of the electronic structure of the QM zone in the following. The wave function of the
electron can be expressed as a linear combination of (orthogonalized) molecular orbitals, $|\phi_m \rangle$, of
fragment molecules A,

\begin{equation}\label{holewavefunc}
\Psi = \sum_{A}\sum_{m \in A} a_m | \phi_m \rangle \;,
\end{equation}
where $\phi_m$ are the molecular orbitals of the individual fragments (molecules). Since low-energy
electron transfer typically occurs in a narrow energy window at the bottom of conduction band, it is sufficient to consider frontier orbitals of
the fragments around the Fermi level\cite{heck2015multi}. Therefore, the set of $\phi_m$ will be restricted to the set of the lowest unoccupied molecular orbitals(LUMOs)
and the total energy expressed in a fragment orbital (FO) basis,

\begin{equation}\label{etot2}
E^- = 
 E^0 - \sum_{A,B} \sum_{m\in A} \sum_{n \in B} a^*_m a_n \langle \phi_m | H[\rho_0] | \phi_n \rangle
 + \Delta E_\mathrm{QM/MM}  \;,
\end{equation}
where $E^0$ is the energy of the charge-neutral system and $H^0_{mn} = \langle \phi_m | H[\rho_0] | \phi_n \rangle$ is the Hamilton
matrix elements, computed using the non-self-consistent variant of the density functional tight-binding
method (DFTB) as discussed in Ref.~\citenum{heck2015multi}. DFTB is derived from DFT by a
Taylor expansion of the total energy around a well-defined reference density, and is found
to be two to three orders of magnitude faster than DFT-GGA functionals with mid-sized basis
sets \cite{elstner2014density,gaus2014density}. Recent studies have shown that the electronic
couplings computed by DFTB are in good agreements with those obtained by other DFT approaches.
Moreover, it is found that application of an uniform scaling factor results in an accuracy
comparable to high-level \textit{ab initio} methods \cite{kubas2015electronic,kubas2014electronic,ziogos2021hab79}.
Therefore, in the present work, the DFTB electronic couplings were scaled by a factor of $1.795$
for electron transport to reach the accuracy of the second-order couple cluster
(CC2) calculations \cite{kubas2015electronic}.

The QM/MM interaction term is given within the DFTB method as
\begin{equation}\label{qmmm}
\Delta E_\mathrm{QM/MM} = \sum_{A} \sum_{m \in A} |a_m|^2
\sum_{K} \sum_{\alpha \in A} \frac{\Delta q^m_{\alpha}q^0_K}
{|R_{\alpha} - R_K|}\;,
\end{equation}
where $q^0_K$ is the partial charge on atom $K$ of the neutral
environment and $\Delta q^m_{\alpha}$ is the change of Mulliken
charge on atom $\alpha$ due to an electron located at orbital $m$.

\subsection{Fewest Switches Surface Hopping}
The propagation equation for the expansion coefficients is obtained by plugging Eq. \ref{holewavefunc}
into time-dependent electronic Schr\"{o}dinger equation (TDSE),
\begin{equation}\label{diabeq}
\dot{a}_m = -i \sum_{n} H_{mn} a_n - \sum_{n} a_n
\langle \phi_m | \dot{\phi}_n \rangle\;,
\end{equation}
where $H_{mn}$ are the DFTB Hamiltonian matrix elements including
the QM/MM interactions. The last term denotes the nonadiabatic coupling
between molecular orbitals $m$ and $n$. This term is negligibly small for molecules far apart ($m \neq n$)
due to the fast exponential decay of molecular orbitals. For $m = n$, the nonadiabatic couplings are
zero as $\langle \phi_m | \dot{\phi}_n \rangle = - \langle \phi_n | \dot{\phi}_m \rangle $ (=0 for m=n).
Therefore, we neglect this term in the integration of the TDSE.

In the FSSH method \cite{tully1990} in the adiabatic representation,
the electronic wave function is expressed as a linear combination of adiabatic basis functions $\{|\psi_i \rangle\}$
\begin{equation}\label{adiabexpan}
  \Psi = \sum C_i^\mathrm{ad} |\psi_i \rangle \;,
\end{equation}
where $C^{ad}_{i}$  are the expansion coefficients of the electronic wave function
in the adiabatic basis set.

Classical trajectories are propagated on a single electronic state according to Newton's equation of motion
\begin{eqnarray}\label{cleom1}
m_K \ddot{R}_k &=& - \frac{\partial H^\mathrm{ad}_j}{\partial R_k} \nonumber \\
&=&  - \frac{\partial  E_\mathrm{MM}^\mathrm{tot}}{\partial R_k} + \sum_{mn} U_{am} U_{an} \left(
\frac{\partial H^0_{mn}}{\partial R_k} - \frac{\partial \Delta E^\mathrm{QM/MM}}{\partial R_k} \right) \;,
\end{eqnarray}
where $H^\mathrm{ad}_a$ denotes the adiabatic energy of the current state $j$. The diagonalization of the electronic
Hamiltonian matrix $\boldsymbol{H}$ (i.e., $-\boldsymbol{H}^{0}+\boldsymbol{H}^{QM/MM}+\boldsymbol{H}^{MM}$)
yields the adiabatic PEs
\begin{equation}\label{diag}
\boldsymbol{H^{ad}} = \boldsymbol{U}\boldsymbol{H}\boldsymbol{U^{\dagger}}\;,
\end{equation}
where $\boldsymbol{U}$ is the diabatic-to-adiabatic (AtD) transformation matrix. Since we only consider the charge
transport in the rigid OSCs with planar backbones where the charge relaxation is much
faster than the charge transfer, it is reasonable to assume the charge
relaxation is instantaneous and neglect the quantum force in Eq. \ref{cleom1}, leading to
\begin{equation}\label{cleomimplicit}
m_K \ddot{R}_k \approx - \frac{\partial E_\mathrm{MM}^\mathrm{tot}}{\partial R_k} \;.
\end{equation}
To consider the effect of charge relaxation on electronic system, the on-site energy, $H_{nn}$
is reduced by a precalculated parameter weighted by charge occupation $\Delta Q_n$ on site
 $n$\cite{kubar2010coarse} (implicit charge relaxation (IR)),
\begin{equation}\label{eq:implicitlambda}
  H^{'}_{nn} = H_{nn} - \lambda_n \Delta Q_n \;.
\end{equation}
Inserting into TDSE, we obtain
\begin{equation}\label{diabeq}
\dot{a}_m = -i \sum_{n} (H_{mn} - \sum_{n} \lambda_n \Delta Q_n \delta_{mn}) a_n \;.
\end{equation}
Note that the precalculated parameter $\lambda_n$ was chosen as
the reorganization energy of a single molecule $n$, computed by $\omega$B97XD and B3LYP, respectively,
with the Def2-TZVP basis set implemented in Gaussian 16 package \cite{gaussian2016}. The charge occupation $\Delta Q_n$ is taken
as wave function population, $|a_n|^2$. The IR approximation avoids the evaluation of the quantum force $\frac{\partial H^0_{mn}}{\partial R_k}$, therefore, reduces
computational cost. Moreover, it allows to remedy the deficiency of DFTB forces by using an
accurate reorganization energy as an input parameter for charge relaxation.

The hopping probability from the current state $j$ to another state $k$ is defined as
\begin{eqnarray}\label{eq:hopprob}
  P_\mathrm{FSSH}^{j \rightarrow k} &=&
   \max \left\{ 0, 2 \mathcal{D}_{jk}
   \frac{\mathrm{Re} \left(C_{k}^\mathrm{ad*} C_{j}^\mathrm{ad}\right)}{|C_{j}^\mathrm{ad} |^{2}} \Delta t \right\}  \;.
\end{eqnarray}
where $\mathcal{D}_{jk}$ is the time derivative coupling between the adiabatic electronic states $j$ and $k$.
A pseudorandom number $\xi$ in [0, 1] is generated and a hopping event from the current state $j$ to state $k$ occurs, if
\begin{equation}\label{eq:hopcondi}
  \sum_{i=1}^{k-1} P_\mathrm{FSSH}^{j \rightarrow i} < \xi \leq
  \sum_{i=1}^{k} P_\mathrm{FSSH}^{j \rightarrow i} \;.
\end{equation}
Note that the traditional velocities adjustment requires the time-consuming computation of the muti-component
the nonadiabatic coupling vector (NCV). In the present work, we introduce an ad hoc approximation,  the Boltzmann correction (BC), by
rescaling the hopping probability with the Boltzmann factor $g_{BC}$ (Boltzmann correction (BC)) ,
\begin{equation}\label{eq:bchopprob}
  P_\mathrm{FSSH}^{j \rightarrow k} \leftrightarrow P_\mathrm{FSSH}^{j \rightarrow k} g_{BC}  \;,
\end{equation}
where
\begin{equation}\label{eq:bc}
g_{BC} =
\begin{cases}
  \exp\left( -\frac{H_{k}^\mathrm{ad} - H_{j}^\mathrm{ad}}{k_\mathrm{B} T}\right), &
       H_{k}^\mathrm{ad} > H_{j}^\mathrm{ad}(\text{upward hops}) \\
  1, & H_{k}^\mathrm{ad} \leq H_{j}^\mathrm{ad}(\text{downward hops}) \;.
\end{cases}
\end{equation}
The velocity remains unchanged during the hopping event. This approximation avoids the computation of
the nonadiabatic coupling vector (NCV) and thus is more efficient.

Note the accuracy of FSSH method using IR and BC approximations has been systematically examined in comparison with the experimental
charge mobilities for a number of OSCs \cite{roosta2022efficient}. For more details, the readers are referred to Ref. \citenum{roosta2022efficient}.

\section{Simulation Details}
The crystal structures of TIPS-TAP , TIPS-TAP-4F , TIPS-TAP-4Cl and TIPS-TAP-4Br were obtained from the cambridge structural database.
The supercells for studied halogen-substituted TAPs contain the following number of molecules along the crystal axis as well as the number of molecules in the QM zone:
105 $\times$ 5 $\times$ 3 (105) for TIPS-TAP, 40 $\times$ 40 $\times$ 3 (36) for TIPS-TAP-4F , 55 $\times$ 55 $\times$ 2 (105) for TIPS-TAP-4Cl,
55 $\times$ 55 $\times$ 2 (104) for TIPS-TAP-4Br. The force field parameters were derived from the general AMBER force field (GAFF)\cite{wang_development_2004,wang_automatic_2006},
where the atomic charges were generated from restrained fitting on the electrostatic potential (RESP) \cite{singh_approach_1984,besler_atomic_1990}
calculated at HF/6-311G* using Gaussian 16 software\cite{gaussian2016}. After an initial minimization, a further NVT simulation of 1 ns at 300 K was
performed to equilibrate the supercell. The V-rescale thermostat \cite{bussi2007canonical} was used to obtain a correct canonical ensemble.
The GROMACS 2018 package was used to perform classical MD simulations \cite{berendsen_gromacs_1995,abraham_gromacs_2015}. The final structure from the equilibrium MD simulation
was taken to start a 10 ns simulation to generate the initial conditions for FSSH-IR-BC simulations. The electron wave function was initially localized on the first molecule,
$\Psi (0) = \phi_1(0)$ and spreads as time increases. A swarm of 500 trajectories was used to obtain converged charge mobilities.
The FSSH-IR-BC method has been implemented in a local version of GROMACS 4.6 simulation package.

The charge carrier mobility is computed with the Einstein-Smoluchowski equation
\begin{equation}\label{mobility}
\mu = \frac{e D}{k_\mathrm{B} T}\;,
\end{equation}
where $k_\mathrm{B}$ is the Boltzmann constant, $e$ is the elementary charge and $T$
denotes temperature. The diffusion coefficient $D$ is calculated by
\begin{equation}\label{diffcons}
D = \frac{1}{2n} \lim_{t \rightarrow \infty} \frac{d \, \mathrm{MSD}(t)}{d t}\;,
\end{equation}
where $n$ is dimensionality ($n=1$ for a one-dimensional chain). The mean square displacement
(MSD)$\mathrm{MSD}(t)$ of the charge is defined as
\begin{equation}\label{msd}
\mathrm{MSD} (t) = \frac{1}{N_\mathrm{traj}}\sum_{l}^{N_\mathrm{traj}} \sum_{A}
(x_A(t)^{(l)} - x_0^{(l)})^2 P(t)^{(l)} (t) \;,
\end{equation}
where $x_A(t)^{(l)}$ and $P (t)^{(l)}$ are the center of mass of molecule $A$
and corresponding charge population along the
trajectory $l$, respectively. $x_0^{(l)}$ is the center of electron at $t = 0$.
The inverse participation ratio (IPR),  which is a measure of the number of molecules
over which electronic wave function is delocalized on average, is defined as
\begin{equation}\label{ipr}
  IPR = \sum_{l=1}^{N_\mathrm{traj}} \frac{1}{\sum_{k}|U_{jk}^{(l)}|^4 }
\end{equation}

\section{Results and Discussion}
\begin{figure}
\begin{center}
\includegraphics[width=0.8\linewidth]{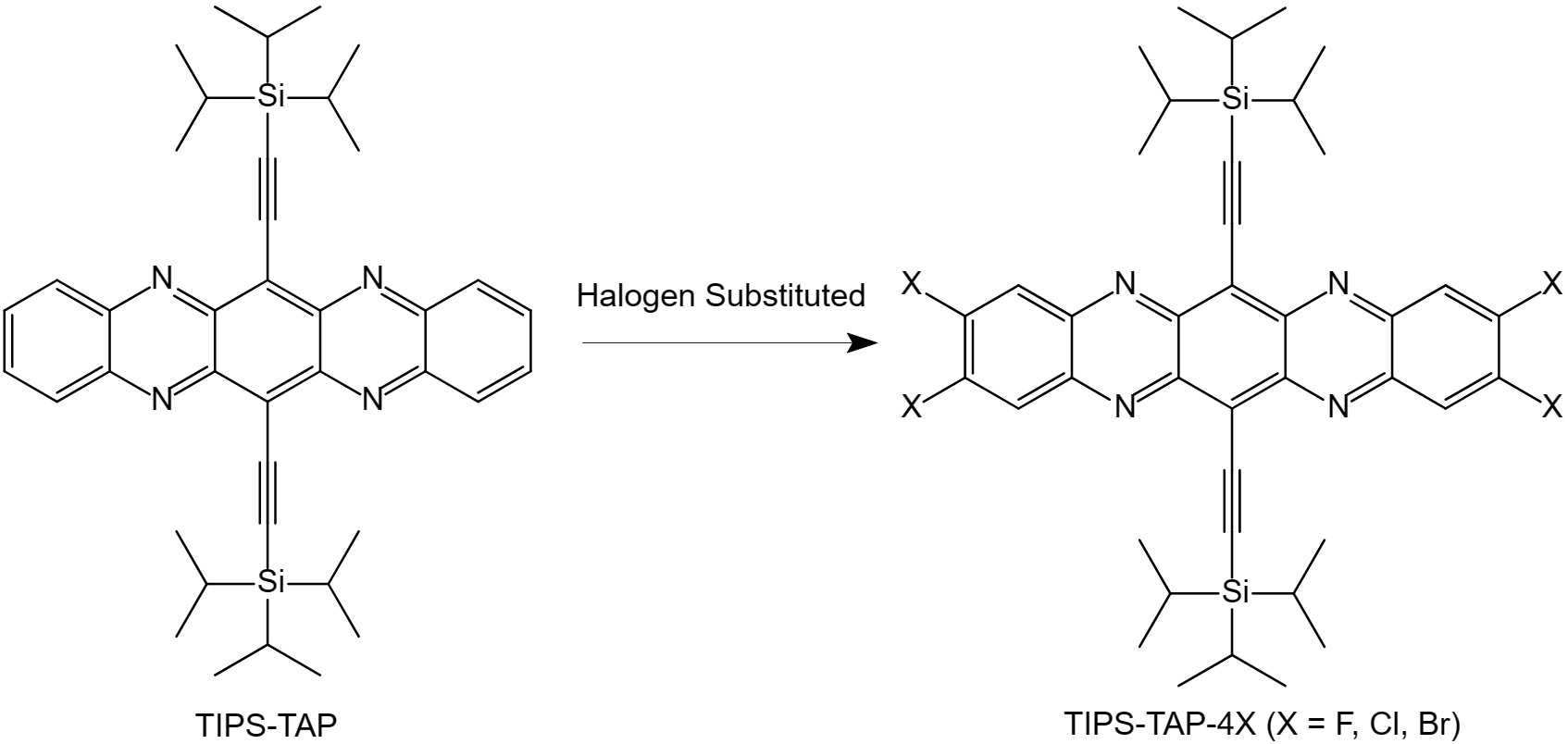}
\vspace{1em}
\caption{Molecular structures of halogen-substituted TIPS-TAP.}
\label{fig1}
\end{center}
\end{figure}

\begin{figure}
\begin{center}
\includegraphics[width=1.0\linewidth]{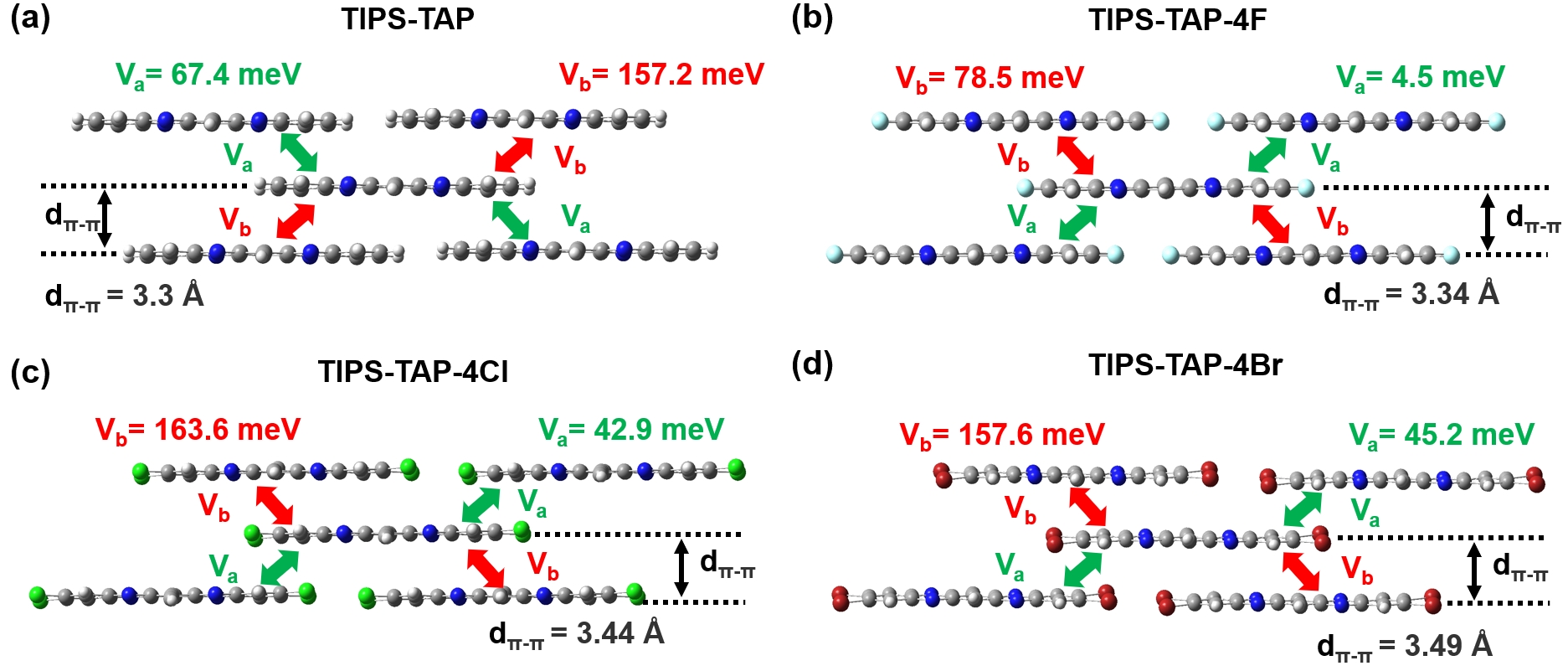}
\vspace{1em}
\caption{Molecular packing of TIPS-TAP and TIPS-TAP-4X (X=F, Cl, Br). Note that the
TIPS groups  have been replaced with hydrogen for clarity. The values of electronic couplings (in meV)
and distances (in $\si{\angstrom}$) of $\pi$-$\pi$ stacking are shown in the figure. }
\label{fig2}
\end{center}
\end{figure}

The molecules considered in the present work are depicted in  Figure \ref{fig1}. Figure \ref{fig2} shows the molecular packing
of TIPS-TAP and its halogen-substituted derivatives. The largest electronic coupling for TIPS-TAP-4Cl  (163.6 meV) and TIPS-TAP-4Br 157.6 meV) are very close to that
for TIPS-TAP (157.2 meV). Interestingly, despite the similar packing structures and close $\pi-\pi$ stacking distances, the electronic coupling for
TIPS-TAP-4F (78.5 meV) is about half of those for TIPS-TAP, TIPS-TAP-4Cl and TIPS-TAP-4Br.
This discrepancy can be attributed to the strong electron-withdrawing effect of fluorine atoms, which results in a more localized LUMO.
To confirm electronic effect of fluorine atoms on the electronic couplings, we calculated the electronic couplings for all halogen-substituted materials,
where the halogen atoms were placed in the MM part and the carbons are connected with hydrogen atoms.
The results are shown in Table \ref{t1}. Upon replacing the chlorine and bromine atoms with hydrogen atoms, we find
that the electronic couplings ($V_{MM}$) remained similar to the original electronic couplings for TIPS-TAP-4Cl and TIPS-TAP-4Br.
However, in the case of TIPS-TAP-4F, replacing the fluorine atoms with hydrogen atoms leads to a substantial increases in the electronic
coupling to 149.7 meV), which is very close to the electronic couplings for TIPS-TAP, TIPS-TAP-4Cl and TIPS-TAP-4Br. This finding
indicates that the electronic effects upon substitution can have a significant influence on the charge transport properties.
Table \ref{t1} summarizes the reorganization energies computed by different methods for all halogen-substituted TIPS-TAPs.
For all materials,  DFTB underestimates the reorganization energies compared to DFT methods (B3LYP and $\omega$B97XD).
The $\omega$B97XD functional gives larger reorganization energies in comparison with B3LYP. Reorganization energies follow the trend:
TIPS-TAP-4F > TIPS-TAP $\sim$ TIPS-TAP-4Cl > TIPS-TAP-4Br.

\begin{figure}
\begin{center}
\includegraphics[width=1.0\linewidth]{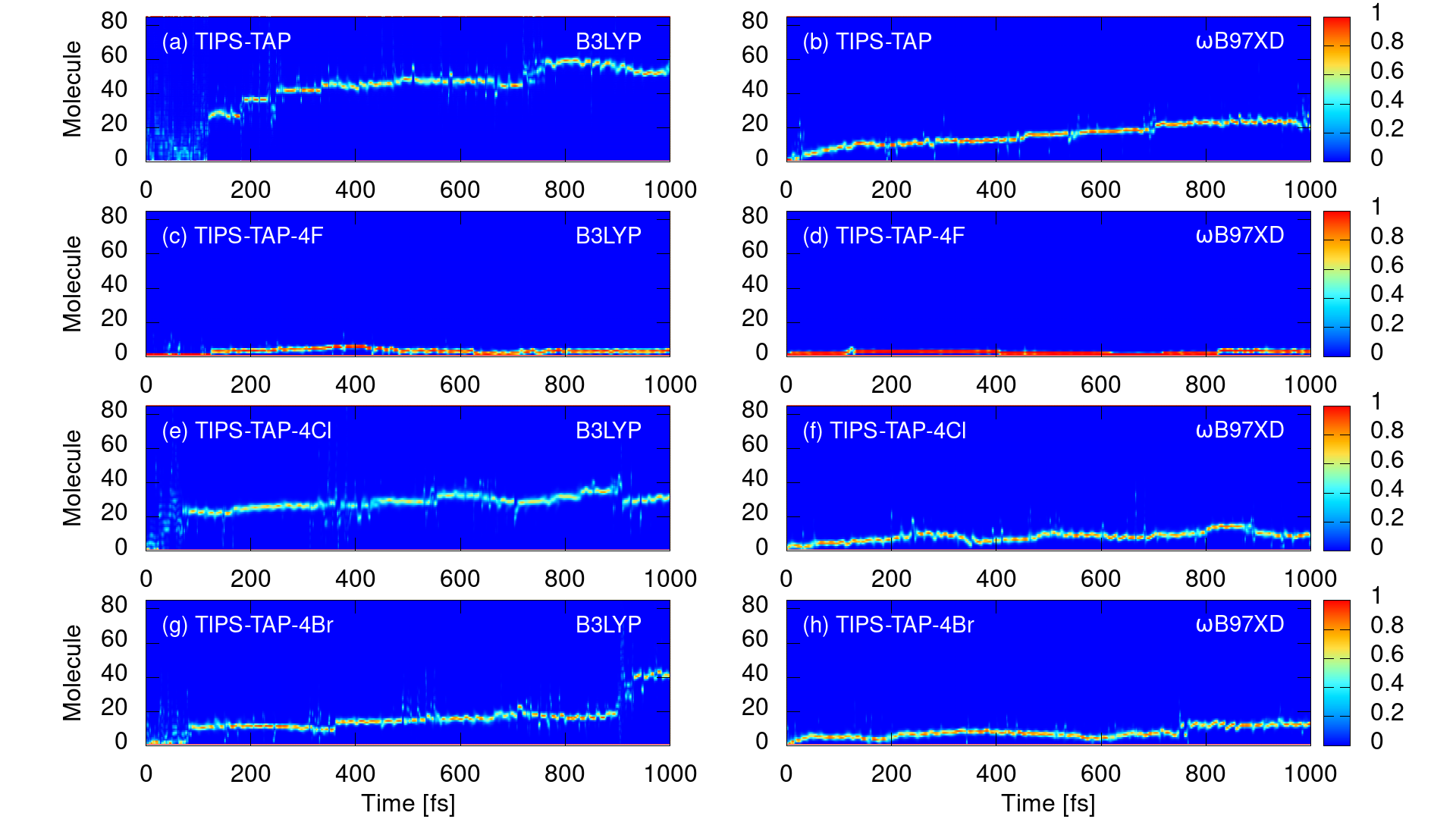}
\vspace{1em}
\caption{Time-dependent electron carrier population for TIPS-TAP and TIPS-TAP-4X (X=F, Cl, Br). The representative
trajectories are propagated by the FSSH-IR method using the reorganization energies calculated
by B3LYP and $\omega$B97XD functionals, respectively.  }
\label{fig3}
\end{center}
\end{figure}

Since the electronic coupling along a direction is significantly weaker than that along b direction,
we performed the FSSH-IR-BC simulations for TIPS-TAP and TIPS-TAP-4X (X=F,Cl,Br) along b direction.
Some representative trajectories propagated with the FSSH-IR-BC method
are shown in Figure \ref{fig3}. Overall,  the electron carrier wavefunction obtained by FSSH-IR-BC using B3LYP
reorganization energy moves faster than that using $\omega$B97XD reorganization energy. For TIPS-TAP, the wavefunction
for  FSSH-IR-BC using B3LYP reorganization energy initially spreads around 20 sites and relocalized on 4-5 molecules after about 100 fs.
This can also be found for TIPS-TAP-4Cl and TIPS-TAP-4Br. However, due to the larger reorganization energies, the fast initial delocalization
is not observed for the FSSH-IR-BC simulations using $\omega$B97XD reorganization energies.
It is evident that the electron carrier in TIPS-TAP-4F is localized onto 2-3 molecules and propagates much slower than other materials.
This can be explained by the large reorganization energy and weak electronic coupling, leading to a slow charge transfer rate.

The comparison of computed electron mobilities and experimental mobilities are shown in Table \ref{t2}.
Overall, the FSSH-IR-BC simulations using B3LYP reorganization energies give lower charge mobilities compared to
the simulations using $\omega$B97XD reorganization energies. For TIPS-TAP, the FSSH-IR-BC simulations using both B3LYP and $\omega$B97XD reorganization energies overestimate
the experimental electron mobility. For TIPS-TAP-4F, the computed mobilities using B3LYP reorganization energies
are in very good agreement with experimental values. This is consistent with the finding of our previous study\cite{roosta2022efficient}
that the FSSH-IR-BC simulations employing B3LYP reorganization energy can reproduce experimental charge mobilities relatively well.
A benchmark study has shown that $\omega$B97XD method gives reorganization energies in excellent agreement with the reorganization energies obtained
by high-level \textit{ab initio} methods \cite{brueckner2016theoretical}. The underestimation of reorganization energy by B3LYP method
leads to a lower barrier height for charge transfer. This effectively compensates for the neglect of the nuclear quantum effects in FSSH simulations and, thus resulting
in a good agreement with experiment. The B3LYP mobility for TIPS-TAP-4Cl coincides with experimental mobility very well.
IPR is a widely used quantity to measure the number of molecules over which the carrier wavefunction is
delocalized. The lower charge mobility indicates the more localized charge carrier. It is found that the trend of IPR is consistent with the mobility change,
where the charge wavefunction of TIPS-TAP-4F is more localized than the other materials.

\section{Conclusions}
In this study, we performed electron propagation using our recently developed FSSH-IR-BC method for a class of n-type organic semiconductors based on halogenated tetraazapentacenes.
The accuracy of FSSH-IR-BC method using the reorganization energies computed by B3LYP and $\omega$B97XD was evaluated by a direct comparison with experimental charge mobilities.
Our results demonstrated good agreement between experiments and the FSSH-IR-BC simulations using B3LYP reorganization energies for the fluorine- and chlorine-substituted
TIPS-TAPs. This agreement can be attributed to the underestimation of reorganization energy, which effectively compensates for the loss of nuclear quantum effects in FSSH simulations\cite{roosta2022efficient},
thus leading to a good agreement with experiment. Interestingly, despite TIPS-TAP-4F exhibiting a molecular packing structure similar to other halogen-substituted TIPS-TAPs, it shows a significantly
lower electron mobility compared to the other materials. Furthermore, when replacing the halogen atoms with hydrogen atoms, TIPS-TAP-4F shows a significant increase in
electronic coupling, whereas the electronic couplings of other materials remain negligibly changed. This finding may be due to the strong electron-withdrawing property of fluorine atoms,
which leads to a weak electronic coupling. While molecular packing is widely regarded as a dominant factor influencing charge transport, our study highlights the crucial role of
electronic effects in charge transport in OSCs.

\section{ACKNOWLEDGEMENTS}
S. R. and M.E. gratefully acknowledge support by the German Research Foundation (DFG) through
SFB 1249 ”N-Heteropolycycles as Functional Materials” (Project B02) and the RTG 2450.
W.X thanks the support from by the National Natural Science Foundation of China (No. 22203047).
All authors acknowledge support by the state of Baden-W\"{u}rttemberg through bwHPC and the
German Research Foundation (DFG) through grant no. INST 40/575-1 FUGG (JUSTUS2 cluster).

\bibliography{./References}

\pagebreak

\pagebreak

\pagebreak

\begin{table}
\caption{Reorganization energies  (in meV), electronic couplings (in meV) and $\pi$-$\pi$ stacking distances (in $\si{\angstrom}$) for TIPS-TAP and TIPS-TAP-4X (X=F, Cl, Br).}
\begin{tabular}{ccccccccccc}
  Crystal  & $d({\si{\angstrom}})$ & $\lambda_{DFTB}$ & $\lambda_{B3LYP}$ & $\lambda_{\omega B97XD}$ &$ <V>$ & $\sigma_{V}$ & <$V_{MM}$> & $\sigma_{V_{MM}}$\\
\hline
TIPS-TAP(a)    &   3.3  & 110  &  205 & 325 & 67.4 &   28.2&- & \\
TIPS-TAP(b)    &  3.3 & 110  &  205 & 325 &157.2 &  29.6 & -& \\
TIPS-TAP-4F(a)    &  3.34  & 122 &  246  & 367 & 4.5  & 33.9 &- &- & \\
TIPS-TAP-4F(b)    &   3.34 & 122 &  246  & 367 &78.5  & 46.0 & 149.7 & 33.9\\
TIPS-TAP-4Cl(a) & 3.44 & 113 &  203  & 318 &42.9 & 15.0 &- &- & \\
TIPS-TAP-4Cl(b) & 3.44 & 113 &  203  & 318 &163.6 & 26.5 & 166.5 & 32.0 \\
TIPS-TAP-4Br(a)   &  3.49 & 111  &  190 & 303 & 45.2  & 12.1 &  -& - &\\
TIPS-TAP-4Br(b)   &   3.49  & 111  &  190 & 303 & 157.6 &  25.0 & 163.6 &  28.8\\
\end{tabular}
\label{t1}
\end{table}

\pagebreak

\pagebreak

\begin{table}
\caption{Computed and experimental electron carrier mobilities (in cm$^2$~V$^{-1}$~s$^{-1}$ ) for TIPS-TAP and TIPS-TAP-4X (X=F, Cl, Br).}
\begin{tabular}{cccccc}
 Crystal  &$\mu_{B3LYP}$  & $\mu_{\omega B97XD}$& $\mu_{Exp}$ & $IPR_{B3LYP}$ &$ IPR_{\omega B97XD}$ \\
\hline
TIPS-TAP(b)   &44.8 &30.6 & 13.3 & 4.5 &3.3\\
TIPS-TAP-4F(b)     &5.5 & 1.9 & 6.6& 2.3& 1.6\\
TIPS-TAP-4Cl(b)  & 27.3& 20.1 & 27.8 & 5.3 & 4.0\\
TIPS-TAP-4Br(b)  &51.3 & 21.5 & - & 4.6 & 3.5\\
          \\
\end{tabular}
\label{t2}
\end{table}

\end{document}